# A study of the quantum classical crossover in the spin dynamics of the 2D *S*=5/2 antiferromagnet Rb$_2$MnF$_4$: neutron scattering, computer simulations, and analytic theories

T Huberman[1], D A Tennant[1,2,3], R A Cowley[1,*], R Coldea[1] and C D Frost[2]

[1]Oxford Physics, Clarendon Laboratory, Parks Road, Oxford, OX1 3PU, UK
[2]ISIS Facility, Rutherford Appleton Laboratory, Chilton, OX15 0QX, UK
[3]Hahn Meitner Institute, Glienicker Str. 100. Berlin D-14109, Germany

**Abstract**
We report comprehensive inelastic neutron scattering measurements of the magnetic excitations in the 2D spin-5/2 Heisenberg antiferromagnet Rb$_2$MnF$_4$ as a function of temperature from deep in the Néel ordered phase up to paramagnetic, $0.13 < k_BT/4JS < 1.4$. Well defined spin-waves are found for wave-vectors larger than the inverse correlation length $\xi^{-1}$ for temperatures up to near the Curie-Weiss temperature, $\Theta_{CW}$. For wave-vectors smaller than $\xi^{-1}$, relaxational dynamics occurs. The observed renormalization of spin-wave energies, and evolution of excitation line-shapes, with increasing temperature are quantitatively compared with finite-temperature spin-wave theory, and computer simulations for classical spins. Random phase approximation calculations provide a good description of the low-temperature renormalisation of spin-waves. In contrast, lifetime broadening calculated using the first Born approximation shows, at best, modest agreement around the zone boundary at low temperatures. Classical dynamics simulations using an appropriate quantum-classical correspondence were found to provide a good description of the intermediate- and high-temperature regimes over all wave-vector and energy scales, and the crossover from quantum to classical dynamics observed around $\Theta_{CW}/S$, where the spin S=5/2. A characterisation of the data over the whole wave-vector/energy/temperature parameter space is given. In this, T$^2$ behaviour is found to dominate the wave-vector and temperature dependence of the line widths over a large parameter range, and no evidence of hydrodynamic behaviour or dynamical scaling behaviour found within the accuracy of the data sets. An efficient and easily implemented classical dynamics methodology is presented that provides a practical method for modelling other semi-classical quantum magnets.

1. Introduction

Powerful inelastic neutron scattering methods are opening up the comprehensive study of dynamics in quantum and molecular magnets. Such quantum spin systems are difficult to model by computer simulations because of the large number of states that have to be included especially for systems in 2 or 3 dimensions. For this reason these systems are frequently modelled by classical spin models for which computations and analytic work is much easier than for quantum systems. The problem is highlighted by considering a spin ½ system for which the angular momentum along the z axis can take only two values ± 1/2, whereas for a

[*] Corresponding author : email address r.cowley1@physics.ox.ac.uk

Quantum classical crossover in the spin dynamics of a 2D antiferromagnet classical description the spin components have a range of possible values depending on the particular angular orientation of the spin. From this simple description it is clear that the classical description of a spin system becomes closer to the quantum system as the spin S becomes large. Further, thermal fluctuations are commonly thought to decrease the importance of quantum effects implying that a crossover to classical behaviour may be observed.

This paper sets out to discuss the use of classical models and analytic theories to describe the spin excitations, over a wide range of temperatures, for a system which is intermediate between those for which a classical spin model is clearly inadequate, and those with extremely large spin, for which it is expected that the classical spin model gives a valid description (this intermediate case is indeed one commonly encountered). We have chosen to study the spin excitations in $Rb_2MnF_4$ using neutron scattering techniques, covering a wide range of temperatures, and comparing the results with analytic theories and classical simulations to determine the extent to which these theories are accurate. $Rb_2MnF_4$ is an approximate realisation of a two-dimensional quantum Heisenberg antiferromagnet on a square lattice, 2DQHAFSL, with a spin value of $S = 5/2$. As the 2DQHAFSL is a central model in statistical physics, the comprehensive mapping of the dynamics presented here is also of broader interest.

Both the structure of $Rb_2MnF_4$ [1], its low temperature excitations [2] and phase transition properties [3] have been studied in detail. The Mermin-Wagner theorem [4] shows that for an ideal two-dimensional system there should not be any long range order at finite temperature if the interactions have isotropic Heisenberg symmetry. In $Rb_2MnF_4$ there is a phase transition to an ordered antiferromagnetic state because of weak exchange constants between the atomic planes and dipolar interactions. Nevertheless the phase transition is at a much lower temperature, by 47%, than expected by mean field theory and so we should expect fluctuations to be important over a wide temperature range. $Rb_2MnF_4$ is then an excellent material to test classical theories because it is intermediate between the strongly quantum spin ½ systems in low dimensional environments and strongly classical systems with large spin and three dimensional environments.

The theory of the excitations in the ideal isotropic 2DQHAFSL, has attracted considerable attention in recent years [5]. Long range order is destroyed above absolute zero and the correlation length then decreases as $\xi = C \exp(A/T)$. Within a correlated region the excitations are expected to be well defined for wavelengths shorter than $\xi$, but for longer wavelengths the excitations will become over damped. Because the model is also relevant for high temperature superconductors, it has been treated with several different approaches. The quantum non-linear sigma model, QNLσM, was developed by Haldane [6] but his approach required $1/S$ to be small. An alternative approach was developed by Chakravarty et al. (CHN) [7] who used symmetry arguments to map the partition function of the Heisenberg model onto the QNLσM in the continuum limit. The model has a quantum critical point with three phases corresponding to quantum disordered, quantum critical and renormalized classical phases. All real systems are believed to be in the renormalized classical part of the phase diagram. In this case there are unsolved questions about whether hydrodynamics and dynamical scaling are valid, even though detailed expressions have been found for the static properties such as the correlation length, as correctly detailed by Hasenfratz [8]. At higher temperatures the properties have been calculated using a variety of techniques and the most successful is possibly the use of the semi-classical self-consistent harmonic approximation, PQSCHA [9] which together with the QNLσM at low temperatures gives a good account of the statics for most temperatures and spin values.

Less success has been achieved in the study of the excitations: Using semi-phenomenological arguments CHN [7] conjecture that the QNLσM exhibits dynamic scaling. They further argue

Quantum classical crossover in the spin dynamics of a 2D antiferromagnet that at low temperatures the excitations should be related to the excitations of the classical rotor model. Quantum Monte Carlo simulations [10], classical Monte Carlo molecular dynamics simulations [11] (similar to the ones described in section 3 and the appendix), and neutron scattering on $Rb_2MnF_4$ [12] (taken with an applied field so as to reduce the effect of anisotropy) all gave results consistent with scaling but with a different temperature dependence for the scaled frequency from that predicted. For $S=1/2$ Ronnow *et al.* [13] have shown that quantum Monte Carlo simulations can be used to give a reasonable description of the dynamical properties of the $S=1/2$ 2DQHAFSL.

A difficulty arises in applying these theories to $Rb_2MnF_4$ because the material is not an ideal 2DQHAFSL and as a result the dipolar anisotropy gives long range order below 38.4K, as described above. This temperature has an energy scale which is almost one half that of the excitations at the zone boundary. It is then not possible to use a continuum model to describe the results in the thermally disordered phase because the high density of zone boundary excitations will dominate the properties and these are not correctly included in a continuum theory. Consequently we have used lattice methods for our calculations. The effect of dipolar anisotropy is considered in detail and we argue that the results when properly averaged over the spin components conform closely to that of the ideal 2DQHAFSL system in the wave vector, energy, temperature range studied.

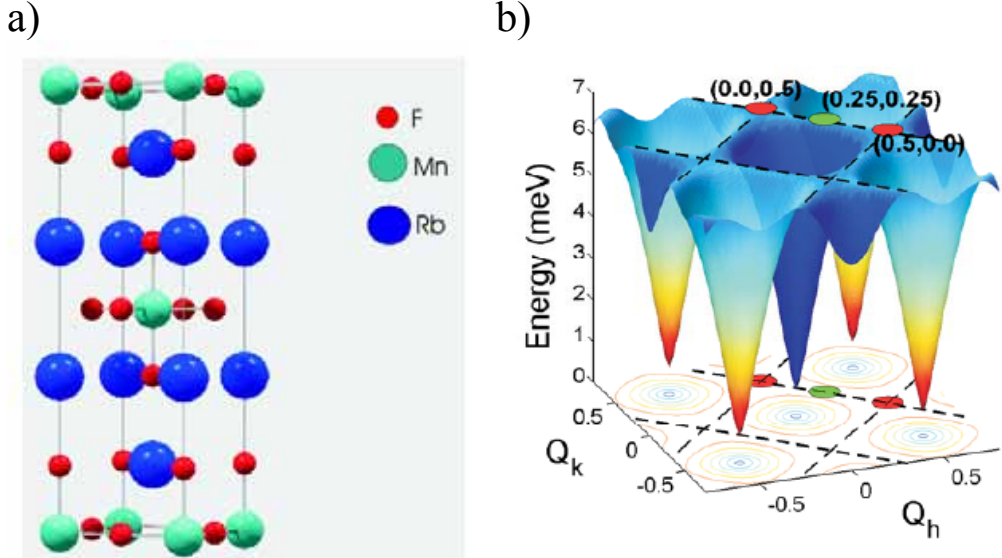

Figure 1 a) The crystal structure of $Rb_2MnF_4$ consists of layers of magnetic $Mn^{2+}$ ions on a square lattice. These are staggered between layers causing frustration, this and the much reduced super-exchange between layers accounts for the two-dimensional magnetism in this material. b) The $T=0$ spin wave dispersion surface as a function of two-dimensional wavevector ($Q_h,Q_k$) and energy $\hbar\omega$. Colour shading is the intensity in neutron scattering with red stronger and blue weaker. Dashed lines in the basal plane and at maximum energy $\hbar\omega = 4J_C Z_C S$ indicate antiferromagnetic zone boundaries. The basal plane also shows equal energy contours (solid lines).

Our experimental results were obtained using the ISIS neutron scattering facility and in section 2 we describe how the experiments were performed and the advantages and disadvantages of this approach. In section 3 we describe analytic theories based on spin waves which are valid at low temperatures in the ordered phase. A random phase approximation is described that gives the temperature dependence of the frequency of the excitations. Calculations are also described for the spin wave lifetime. At higher temperatures and in particular above the transition temperature the spin wave approaches are possibly invalid and we then use results obtained by classical Monte Carlo techniques to calculate the scattering in section 4. The results of the neutron scattering measurements, the analytic

Quantum classical crossover in the spin dynamics of a 2D antiferromagnet theories and the computer simulations are then compared within section 5, and finally the results of our study discussed in section 6. A comprehensive appendix covering the classical dynamics computations is presented at the end.

## 2. Rb$_2$MnF$_4$ and Neutron Scattering

Rb$_2$MnF$_4$ is a good antiferromagnetic insulator with a Curie-Weiss temperature $\Theta_{CW}$=87.5K. It crystallizes in the same tetragonal K$_2$NiF$_4$ structure as the high temperature superconductor parent compound La$_2$CuO$_4$, with space group *I4/mmm* and lattice parameters *a=b*=4.215 Å, and *c*=13.77 Å, see fig. 1. The magnetic properties are largely two-dimensional with planes of Mn$^{2+}$ ions carrying a spin-only moment of S=5/2 arranged on a square lattice in the crystal *a-b* plane. Strong antiferrromagnetic superexchange interactions occur between nearest neighbour Mn$^{2+}$ ions in each plane through the intervening F$^-$ ions. The coupling between planes along the *c*-direction is in comparison some 10$^{-4}$ times weaker due to the magnetic MnF$_2$ planes being separated by two nonmagnetic RbF layers. In addition the successive MnF$_2$ layers are staggered such that Mn ions lie above the centre of the Mn-F plaquettes and so are equidistant to the four Mn atoms in the layer below (and above). The wave vectors are written in reduced form (Q$_h$,Q$_k$,Q$_l$) where the units are 2π/*a*, 2π/*a* and 2π/*c*. Cartesian coordinates $\hat{\mathbf{x}}, \hat{\mathbf{y}}, \hat{\mathbf{z}}$, as applied to the spin components, are parallel to the crystal **a**, **b**, and **c** directions respectively.

The magnetic structure of Rb$_2$MnF$_4$ was studied by Birgeneau *et al.* [1] and it was found that below a temperature $T_N$ =38.4K a staggered sub-lattice magnetisation developed with the spins aligned perpendicular to the magnetic planes. There was a much weaker tendency to order between the planes and the results depended on the rate of cooling and other details. The magnetic excitations were first measured by Cowley *et al.* [2], who found that at low temperatures there were well defined spin waves that could be largely explained in terms of a spin Hamiltonian:

$$H = J\sum_{i,i'}(S_i^x S_{i'}^x + S_i^y S_{i'}^y + S_i^z S_{i'}^z(1+\Delta)) \quad (1)$$

with exchange strength J=0.6544(14) meV. In this equation the summation is restricted to nearest neighbours and the anisotropy term, Δ=0.0048(10), probably arises from the dipole-dipole interactions. The experiments showed that there was also a very much weaker next nearest neighbour interaction. This has been studied in more detail in our recent paper [14] and has strength *J'*=0.006(3) meV, but we shall not consider it further as it is 10$^{-2}$ times weaker than the nearest neighbour exchange. In these measurements the high accuracy of spin wave theory was confirmed at low temperatures, both in the energy and intensity of the one magnon excitations in Rb$_2$MnF$_4$, and also in the two-magnon scattering. Interactions at lowest temperatures for the spin waves actually renormalise the excitation spectrum by a small factor $Z_C$=(1+0.157/(2S))=1.0314. This factor has been absorbed in the exchange strength $J=Z_C J_C$=0.6544(14) meV where the true exchange strength of the material is $J_C$=0.6345(14) meV. In comparing between quantum and classical models later, the quantum renormalisation $Z_C$ will be found to be small compared to the apparent renormalisation in spin length between low and high temperatures, $\sqrt{S(S+1)}/S = 1.183$. This property makes Rb$_2$MnF$_4$ an ideal system in which to detail such a crossover.

Dynamical neutron scattering measures the Fourier transformed pair correlation functions:

$$S^{\alpha\beta}(\mathbf{Q},\omega) = \frac{1}{2\pi N}\sum_{i,i'} e^{i\mathbf{Q}\cdot(\mathbf{R}_i-\mathbf{R}_{i'})} \int_{-\infty}^{\infty} e^{-i\omega t} \langle S_i^\alpha(t_0) S_{i'}^\beta(t_0+t)\rangle dt.$$

For a Heisenberg type exchange the off-diagonal components are identically zero, *i.e.* $S^{\alpha\beta}(\mathbf{Q},\omega) = 0$ when $\alpha \neq \beta$, and in addition by symmetry $S^{xx}(\mathbf{Q},\omega) = S^{yy}(\mathbf{Q},\omega) \neq S^{zz}(\mathbf{Q},\omega)$ due to the small anisotropy. To measure the dynamical correlations, experiments were performed on a 13.4 g single crystal of Rb$_2$MnF$_4$ using the same methodology as in [14]. The

Quantum classical crossover in the spin dynamics of a 2D antiferromagnet

MAPS instrument [15] at the ISIS facility of the Rutherford Appleton Laboratory was used. MAPS is a direct geometry time-of-flight neutron scattering spectrometer and is equipped with a 16m$^2$ array of position sensitive detectors which are divided into nearly 40,000 different elements. The sample was placed in a cryostat that enabled the temperature to be varied between 9 K and 300 K, and was aligned so that the c-axis, perpendicular to the two-dimensional magnetic planes, was horizontal and parallel to the incident beam of neutrons, and with a crystallographic [100] axis horizontal and perpendicular to the incident beam. Data was collected with an incident neutron energy of $E_i$=24.92 meV using a chopper speed of 300 Hz. This enabled the detectors at low angles to collect scattering from around all four of the lowest angle magnetic Bragg reflections. Because the scattering is, in principle, the same around each of these Bragg reflections the results were added together to improve the statistics. The data is a highly pixelated 3D volume in the 4D ($Q_h$, $Q_k$, $Q_l$, $\hbar\omega$) space. As there is little correlation between the magnetic planes, and the scattering depends on the components of the in-plane wave vector $Q_h$, $Q_k$ and the energy transfer, $\hbar\omega$, the $Q_l$ component is projected out and a compact 3D volume in ($Q_h$, $Q_k$, $\hbar\omega$) space rendered. The scattering intensities are corrected for the magnetic form factor of Mn$^{2+}$, making it proportional to the scattering law $S(\mathbf{Q}, \omega)$ convolved with the instrumental resolution. Magnetic neutron scattering probes those spin components perpendicular to the wave vector transfer $\mathbf{Q}$, so $S(\mathbf{Q},\omega) = (2-p_z)S^{xx}(\mathbf{Q},\omega) + p_z S^{zz}(\mathbf{Q},\omega)$ where the polarization factor $p_z = 1-\hat{Q}_z^2$ and $\hat{Q}_z$ is the directional cosine along the c-direction of the wavevector $\mathbf{Q}$. Spinwaves have a transverse character and are seen in the xx and yy components in the ordered phase. MAPS effectively integrates over the spin components.

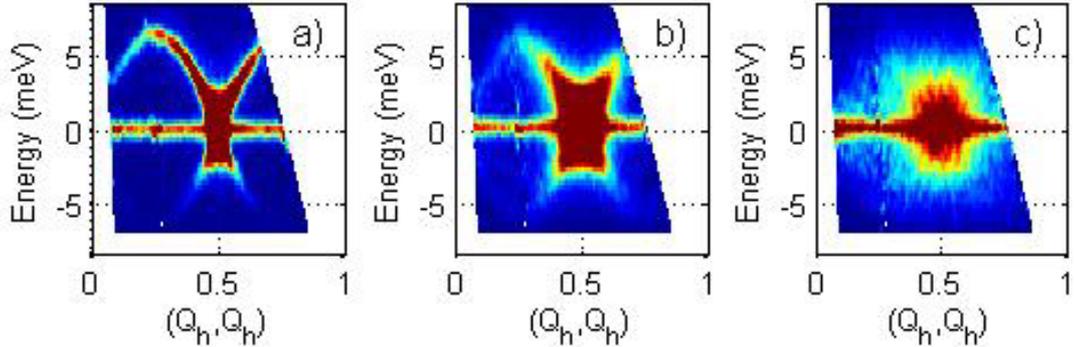

Figure 2. Cross sectional slices through the 3D volumes in ($Q_h$, $Q_k$, $\hbar\omega$) space rendered from the MAPS data. Colour shading is the intensity in neutron scattering with red stronger and blue weaker. Slices are shown for temperatures a) 21.3 K which is well below $T_N$, in the three-dimensionally ordered phase. Well defined transverse spin waves are observed across the Brillouin zone. b) 46.9 K, above $T_N$ but with significant short range order. Spin waves are observed although with increased lifetime broadening. The spin waves are over-damped for wave lengths longer than the correlation length. c) 100.7 K, above the Curie Weiss temperature. Over-damped behaviour alone is seen in the paramagnetic phase. Note that these slices are taken from the measured data. No background has been subtracted and in particular incoherent scattering is seen at zero energy transfer.

Representative data is shown as a series of coloured plots in fig. 2, with the colour indicating the intensity of scattering. The results are qualitatively as expected in that as the temperature is raised 1) the energy of the excitations decreases, and 2) the line-width of the excitations increases. To aid a detailed comparison of the experiment with theory we have chosen to display mostly plots of the data taken with a constant wave vector $\mathbf{Q}$. The data is available for $0 < Q_h < 0.5$ and $0 < Q_k < 0.5$ in units of $2\pi/a$ and will be plotted from (0,0) along the direction (q,0), along the Brillouin zone boundary (q, 0.5-q), and through the antiferromagnetic reciprocal lattice point (q,0.5) with q varying from 0 to 0.5.

Quantum classical crossover in the spin dynamics of a 2D antiferromagnet

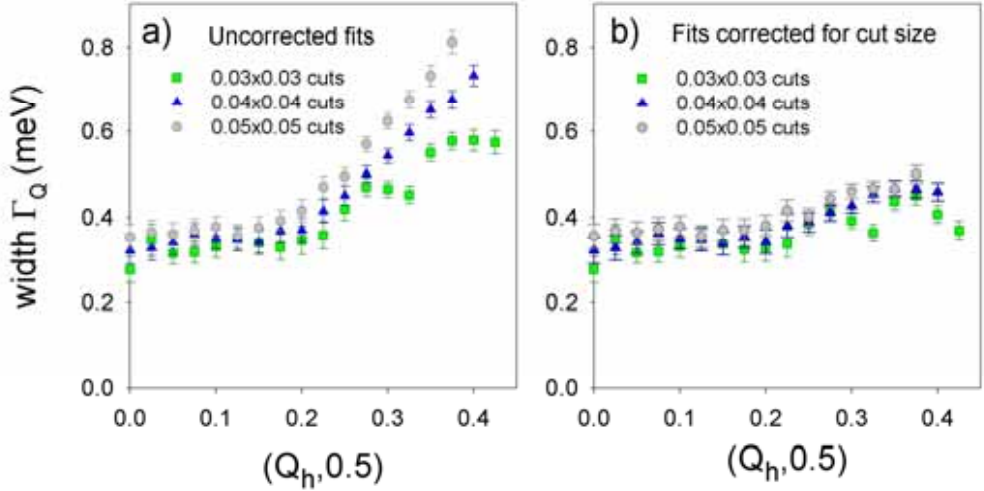

Figure 3. The widths of the excitations at a temperature of 25.5K. The widths have been extracted using three different areas for summing over for each point (cut sizes): 0.03x0.03, 0.04x0.04 and 0.05x0.05 reciprocal lattice units. a) shows the results if there is no correction for the finite cut size whereas b) includes the correction and there is then little difference between the results.

Although there is no analytical form for the structure factor of the excitations, we find the data to be well described by a Lorentzian line shape:

$$L(Q,\omega) = \frac{1}{\pi(1-\exp(-\hbar\omega/kT))}\left[\frac{\Gamma}{(\hbar\omega-\hbar\omega_Q)^2 + \Gamma^2} - \frac{\Gamma}{(\hbar\omega+\hbar\omega_Q)^2 + \Gamma^2}\right] \quad (2)$$

The widths and frequencies of excitations at each wavevector and temperature can then be extracted and compared between theory and experiment. The analysis of the data from MAPS is furthermore complicated because the experimenter can choose the area over which to sum the original data before displaying it and this plays a similar role to that of the resolution for a conventional neutron instrument. The area to be summed over in wave vector is chosen to be as small as possible so as to have the best possible resolution but large enough to have a good statistical accuracy. The effect of this compromise is illustrated in figure 3a: Constant Q scans (i.e. scans along the energy axis at a wavevector **Q**) were obtained by using areas in wave vector centred on wavevector **Q** of dimension $\Delta Q_h \times \Delta Q_k$ = 0.03 x 0.03, 0.04 x 0.04 and 0.05 x 0.05 in wave vector units of $2\pi/a$ for each point in the scan. Then the linewidth of the spinwave excitations within these scans was fitted. Fig 3a shows the obtained linewidths of these points in wave vector. Clearly the larger the size of the area chosen the wider the linewidth obtained. Because of this the influence of the cut as well as resolution of the instrument was also included into the fitting procedure. The technique used was to consider the scattering if the incident neutron had an energy $\varepsilon_0$ and then to sum over a range of incident energies with probabilities $p(\varepsilon - \varepsilon_0)$. The probability function was then chosen to give the observed widths for both the incoherent elastic width and the low temperature width of the excitations with wave vectors at the zone boundary. Both of these were found to be 0.3 meV and the form of the probability function was taken to be a Gaussian.

The cross-section from an excitation was then assumed to be of a Lorentzian form convoluted with the distribution of the incident energy to give a Voigt function $V(Q,\omega)$. The spectrum was averaged over the cut area $\Delta Q_h \times \Delta Q_k$ in reciprocal space by summing over a discrete set of points to give:

$$I(Q,\omega) = \frac{1}{N}\sum_i \frac{1}{1-\exp(-\hbar\omega/kT)} V(Q_i,\omega) \quad (3)$$

Quantum classical crossover in the spin dynamics of a 2D antiferromagnet where $Q_i$ is summed over several hundred wave vectors within each cut. In the expression for the Voigt function the expression for the spin wave frequencies was used to obtain the derivatives of the frequencies at each wave vector with the exchange constants adjusted so as to obtain a satisfactory fit for each temperature. Figure 3b shows the widths of the excitations when the resolution is included and there is reasonably good agreement between the results for different sizes of the original cuts. This procedure was then used for the analysis of all of the results and gave accurate estimates for the frequency and energy at each temperature.

The temperature dependence of the excitations was studied from low temperature to high temperature with temperatures of 9.5, 21.3, 25.5, 30.2, 35.4 K below the ordering temperature of $T_N$=38.4K, and temperatures 40.6, 46.9, 51.6, 56.7, 62.8, and 100.7 K above the ordering temperature. The data could be rapidly obtained (in about 8 hrs per temperature) simultaneously for the energy and line-width of the excitations over a large part of the Brillouin zone. It is substantially faster than to obtain similar data over as wide an area of reciprocal space using a conventional triple axis machine. Unfortunately the use of a spallation time-of-flight machine inevitably results in a resolution that is not tailored to give the best resolution for each particular wave-vector transfer. In the case of MAPS measurements on Rb$_2$MnF$_4$ the low energy spin waves near the antiferromagnetic lattice point (0.5, 0.5) have worse resolution than the higher energy zone boundary modes, whereas ideally the resolution should be better for the low energy modes. The result of this is also that the excitations at low energies could not be separated easily from the incoherent elastic scattering. The results therefore for the lowest energy spin waves are not as good as would be obtained in a conventional study with a triple axis spectrometer. The time-of-flight technique is, however, very efficient and very well suited for the study of the excitations at the zone boundary. It would probably be unreasonably lengthy to study as many temperatures as we have done here and the detailed behaviour of the zone boundary with a conventional neutron technique.

### 3. Analytic Calculations of Spin Waves
*3.1 Energies*

The conventional theory of spin waves is described in many texts and so we shall not discuss the derivation in detail [16]. It assumes that the ground state has long range antiferromagnetic order, and the result of using the Hamiltonian eqn. 1 with nearest neighbour interactions gives the energy as:

$$\hbar\omega(q) = 4JS(A^2 - B^2\gamma(q)^2)^{1/2} \quad (4)$$

where the wave-vector dependent term is

$$\gamma(q) = \cos(\pi(q_h + q_k))\cos(\pi(q_h - q_k))$$

The terms A and B are given using the random phase approximation by:

$$A = 1 + \Delta - C(T) - \Delta D(T)$$
$$B = 1 - C(T) - \Delta(C(T) - D(T))$$
$$C(T) = \frac{1}{NS}\sum_q (n_q + 1/2)\left[\frac{A - \gamma(q)^2 B}{(A^2 - \gamma(q)^2 B^2)^{1/2}}\right] - \frac{1}{2} \quad (5)$$
$$D(T) = \frac{1}{NS}\sum_q (n_q + 1/2)\left[\frac{A}{(A^2 - \gamma(q)^2 B^2)^{1/2}}\right] - \frac{1}{2}$$

where the two terms, C(T) and D(T), arise from the spin wave interactions, and $n_q$ is the Bose occupation factor. If the interactions with the zero point fluctuations are put to zero, C(T) and D(T) = 0, eqn. 4 gives the normal non-interacting spin wave result for the two-dimensional square lattice [17]. Otherwise these equations for the energies of the excitations can be calculated self-consistently. This gives a quantum solution for the energy of the excitations that has these corrections even at absolute zero. The classical solution is obtained by replacing the spin wave occupation factor ($n_q$+1/2) by $k_B T/\hbar\omega_q$. One clear defect of this RPA

Quantum classical crossover in the spin dynamics of a 2D antiferromagnet approximation is that the modes do not have a finite line-width; the calculation of this is discussed in the next section.

*3.2 Line-widths*

The main damping in the 2DQHAFSL is expected to arise from spin wave-spin wave scattering, which can be calculated using perturbation theory. The theory can be expected to be valid locally within regions much smaller than the correlation length *i.e.* where the reduced wave-vector from the antiferromagnetic point $|q|>>\kappa(T)$ for $T>T_N$, and for all $q$ in the ordered phase (for $T<T_N$), where $\kappa(T)$ is the inverse correlation length of the magnetic order. The lowest Born approximation for the damping is given by Tyč and Halperin [18] as:

$$\Gamma_\mathbf{q} = \frac{\pi}{16S^2(2\pi)^4}(1-\exp(-\hbar\omega(\mathbf{q})/kT))\times \int d^2k d^2p\, n_p(1+n_r)(1+n_s)M_{22}(\mathbf{q},\mathbf{p},\mathbf{r},\mathbf{s})\delta(\Delta\omega)$$ (6)

where **q** and **p** are the wave vectors of the incoming spin waves, and **r** and **s** are the wave vectors of the outgoing spin waves. These satisfy the conservation of momentum $\mathbf{q}+\mathbf{p}=\mathbf{r}+\mathbf{s}$, and energy $\Delta\omega = \omega(\mathbf{q})+\omega(\mathbf{p})-\omega(\mathbf{r})-\omega(\mathbf{s})$. The matrix element $M_{22}$ is that for a two-spin-waves-in to two-spin-waves-out scattering process and has the following form obtained by Tyc and Halperin [18] at long wavelengths:

$$M_{22} = 2\left[(1-\hat{\mathbf{q}}\cdot\hat{\mathbf{p}})(1-\hat{\mathbf{r}}\cdot\hat{\mathbf{s}}) + (1-\hat{\mathbf{q}}\cdot\hat{\mathbf{r}})(1-\hat{\mathbf{p}}\cdot\hat{\mathbf{s}}) + (1-\hat{\mathbf{q}}\cdot\hat{\mathbf{s}})(1-\hat{\mathbf{p}}\cdot\hat{\mathbf{r}})\right].$$

While Tyč and Halperin give analytic forms for various restricted temperature regimes, the line broadening calculated below was evaluated numerically and the approximation was improved by taking the temperature renormalized spin wave dispersion.

Kopietz [19] has described calculations for short wavelengths with a more accurate form for $M_{22}$, valid in the range $|q|>>(2\pi/a)[k_B T a/c]^{1/3}$, where the slope $c = 2^{3/2}JSa = a\hbar\omega_{ZB}/\sqrt{2}$, and the lattice parameter $a$=4.215 Å. His proposed form for the line width is

$$\Gamma_q = \omega_{ZB}\frac{2\pi}{3S^2}\left(\frac{k_B T}{\hbar\omega_{ZB}}\right)^3 Z(|\mathbf{v}(\mathbf{q})|)$$ (7)

Where $Z(|\mathbf{v}(q)|)$ is a numerically computed function [19] and $\mathbf{v}(\mathbf{q})$ is the gradient of the spin wave dispersion at wave vector **q**.

### 4. Classical Simulations

Calculations of the frequency and lifetime of the excitations have not been performed analytically at high temperature and particularly above $T_N$. We have therefore performed numerical simulations using a classical model and a Monte Carlo procedure. The procedure is essentially similar to that developed by Metropolis *et al.* [20] and described in detail by Binder and Heermann [21] and for this specific case by one of us [22] and in the appendix. The energy of the system is treated classically which means that the spins are taken as being of unit length and can point in any direction to produce a spin configuration. This is up-dated by allowing one spin to change its orientation and then considering whether or not the energy of the system has increased or decreased. Unfortunately if single spins are considered the algorithm is slow and so many of the calculations were performed with an over-relaxation algorithm which uses configurations that are as far as possible from the previous configuration. This enables phase space to be covered more efficiently than with single spin up-dates and helps to prevent the system becoming stuck in a false minimum.

The computed transverse response $S^{xx}(Q,\omega)+S^{yy}(Q,\omega)$, perpendicular to the hard axis (z-direction), is shown for a wave vector **Q**=(0.4,0.1) and at four different temperatures in fig. 4,.

Quantum classical crossover in the spin dynamics of a 2D antiferromagnet

There is, as expected, a sharp peak at low temperatures that corresponds to the spin wave response. This decreases in frequency (energy) as the temperature is raised while the line-width instead increases. To extract the frequency and line-width of the excitations from such computed data a modified Lorentzian form was fitted:

$$S_c(Q,\omega) = \frac{\exp(\hbar\omega/2kT)}{\pi(\exp(\hbar\omega/kT)-1)}\left[\frac{\Gamma}{(\hbar\omega-\hbar\omega_Q)^2+\Gamma^2} - \frac{\Gamma}{(\hbar\omega+\hbar\omega_Q)^2+\Gamma^2}\right] \quad (8)$$

This differs from the similar expression that was fitted to the experimental results because the calculations have classical symmetry, and so this expression has also been chosen to satisfy the attendant classical symmetry $S_c(Q,\omega) = S_c(Q,-\omega)$. Note that with the quantum mechanical cross-section detailed balance holds, i.e. $S(\mathbf{Q},-\omega) = e^{-\hbar\omega/k_BT}S(\mathbf{Q},\omega)$ (and of course by reflection symmetry of the system $S(\mathbf{Q},\omega) = S(-\mathbf{Q},\omega)$), and the quantum symmetry can be regained from a classical simulation using:

$$S(\mathbf{Q},\omega) \approx e^{\hbar\omega/2k_BT}S_C(\mathbf{Q},\omega)\left(\int_{-\infty}^{\infty}d\omega S_C(\mathbf{Q},\omega) \Big/ \int_{-\infty}^{\infty}d\omega \cdot e^{\hbar\omega/2k_BT}S_C(\mathbf{Q},\omega)\right), \quad (9)$$

which has been normalised to retain the integrated intensity of the classical cross section. This will be used for the direct comparison with line shape, e.g. in figs. 8 and 9. So comparatively the classical symmetry decreases the intensity on the energy loss side while increasing the intensity on the energy gain side.

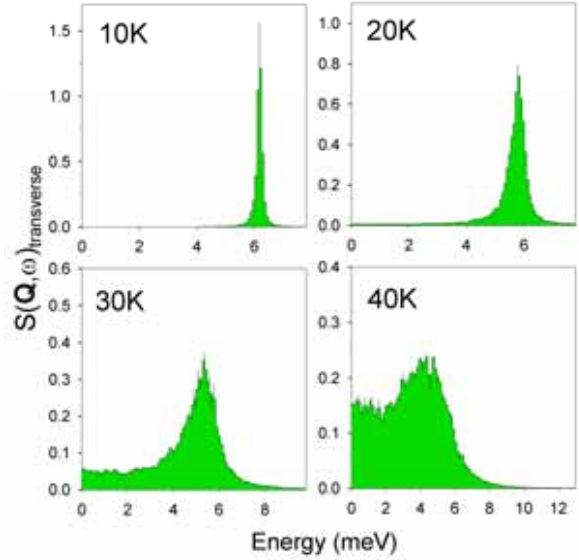

Figure 4 The results of simulations of the transverse spin scattering for several temperatures. The wave vector is (0.4,0.1) reciprocal lattice units.

The longitudinal scattering (along the hard axis) $S^{zz}(Q,\omega)$, was also calculated and the results are shown in fig. 5. The simulations were performed at 10K and there is good agreement between the results and those of two spin-wave theory, solid line, which were calculated in our earlier paper [14]. This agreement shows that the simulations are very satisfactory because the longitudinal cross section is small at low temperatures when the simulations have most difficulty in achieving equilibrium. At higher temperatures, and particularly above the ordering temperature, the shape of the longitudinal scattering is different from two-spin wave scattering and becomes very similar in shape to that of the transverse components of the scattering. However, just above $T_N$ the simulations showed that there was at least short range symmetry breaking because the longitudinal and transverse cross sections were then significantly different. We also showed that at high temperatures the simulated scattering was qualitatively consistent with the expected form of the paramagnetic scattering and conclude

Quantum classical crossover in the spin dynamics of a 2D antiferromagnet that since the simulations are in agreement with calculations at both low and high temperatures they are likely to be correct for the full temperature range.

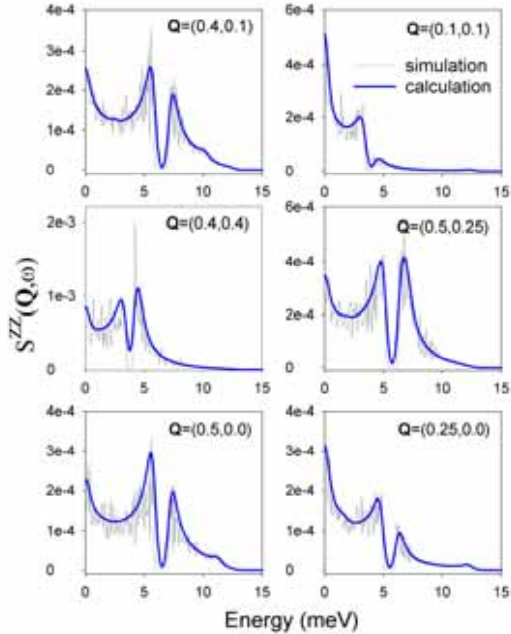

Figure 5 A comparison of the simulated and analytical expression for $S^{zz}(Q,\omega)$ for various wave vectors. The low intensity of the scattering and the good agreement with theory shows that the simulations are very satisfactory.

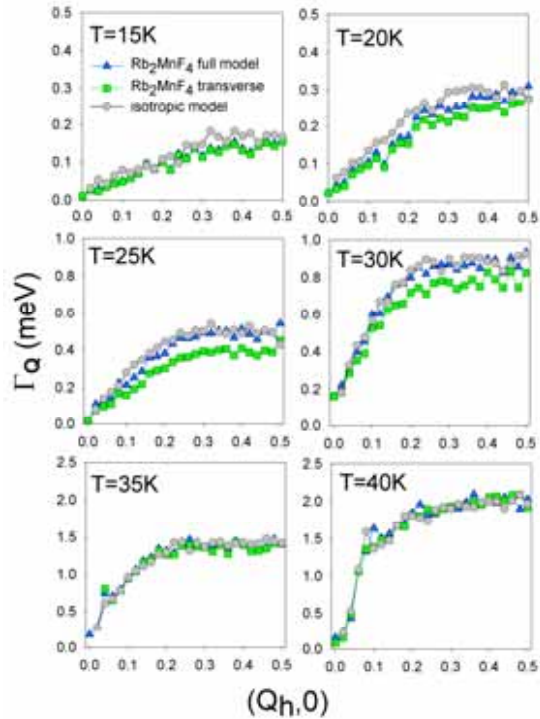

Figure 6 The results of fitting a Lorentzian line shape to the simulation results for $S(Q,\omega)_{isotropic}$ the dynamical cross-section of the isotropic model, $S(Q,\omega)_{crystal}$ which is a computation of the dynamical correlations for $Rb_2MnF_4$ including the anisotropy and for all the *x,y,* and *z* components, and $S(Q,\omega)_{transverse}$ which again is for $Rb_2MnF_4$ including the anisotropy but only including the transverse spin components *xx* and y*y*.

Quantum classical crossover in the spin dynamics of a 2D antiferromagnet

An important question is to what degree the line-widths are modified by the small (0.5%) dipole anisotropy term $\Delta=0.0048$. To ascertain its importance classical simulations were also performed for the isotropic Heisenberg model with $\Delta=0$. Figure 6 shows the line-width of the excitations calculated for 1) the isotropic model *i.e.* $\Delta=0$, 2) that deduced from the transverse part of the scattering from $Rb_2MnF_4$ with $\Delta=0.0048$, and 3) from extracting the line width from $S^{xx}(\mathbf{Q},\omega)+S^{yy}(\mathbf{Q},\omega)+S^{zz}(\mathbf{Q},\omega)$, the total scattering, again simulated for $Rb_2MnF_4$ with $\Delta=0.0048$. There is little difference between the different curves especially between the isotropic model and the averaged total scattering from $Rb_2MnF_4$, especially at temperatures of 35 K and above. At lower temperatures, below 20K, the isotropic model gives better agreement with the total scattering from the $Rb_2MnF_4$ while at the lowest temperatures the isotropic model gives peaks that are slightly wider than the simulations of $Rb_2MnF_4$. The temperature dependence of the frequencies of the isotropic excitations is shown in fig. 7 and the frequencies of the excitations steadily decrease with increasing temperature. The random phase approximation describes the results accurately especially below $T_N$. In conclusion, then, for the temperatures studied here the actual measured line-widths of $Rb_2MnF_4$ should approximate well to the behaviour of the ideal isotropic Heisenberg lattice (the 2DQHAFSL).

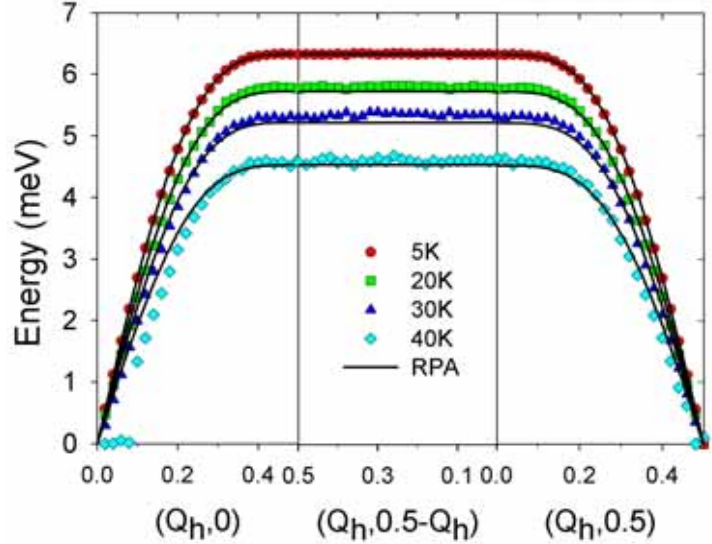

Figure 7 The dispersion relations of the isotropic model obtained from the simulations and the results compared with the random phase approximation below 40 K.

Before a detailed comparison between the experiment and theory is possible we need to choose the parameters that are used for the calculations with those deduced from the experiment using low temperature spin wave theory. In principle, the quantum mechanical Hamiltonian could be replaced by an effective classical Hamiltonian and the parameters of the effective Hamiltonian, $S_c$ and $J$, could be chosen to vary as a function of temperature so as to reproduced the observed results. This general solution is complex and we have preferred to use an effective Hamiltonian for which the parameters are readily deduced from the quantum Hamiltonian at high or low temperature. The integral of the scattering over frequency and wave vector is proportional to $S(S+1)$ for a quantum system and $S_c^2$ for the classical system with spin $S_c$. At high temperatures the integral over the scattering for each wave vector is also proportional to $S(S+1)$ for the quantum system and $S_c^2$ for the classical system. Furthermore the ground state energy is the same apart from the zero point energy of the spin waves for the classical and quantum systems provided that $S_c^2 = S(S+1)$. A difficulty arises, however, when the expressions for the low temperature excitations are considered. For the quantum system the energy of the spin waves is proportional to $S$ with a possible correction term of $(1+0.157/(2S))$ which arises from spin wave interactions at $T = 0$ K, whereas for the classical system the energy of the excitations is proportional to $S_c$. If we use $S_c^2 = S(S+1)$ then the

Quantum classical crossover in the spin dynamics of a 2D antiferromagnet classical calculations will give excitations with an energy that is larger than that deduced from the quantum calculations at low temperatures. This is because a classical calculation neglects the quantum fluctuations and these are particularly important at low temperature. Because of this problem the simulations have also been performed with two models: Model A has $S_c^2 = S(S+1)$ and $J = 0.63$ meV and model B has $S_c = S$ and $J=0.65$ meV. The latter model provides a good fit to the low temperature spin waves whereas the former is expected to describe the high temperature properties of the excitations accurately. Model A has a simple appeal for a comparison to the true quantum behaviour as it uses the actual coupling strength with a straightforward spin length correspondence.

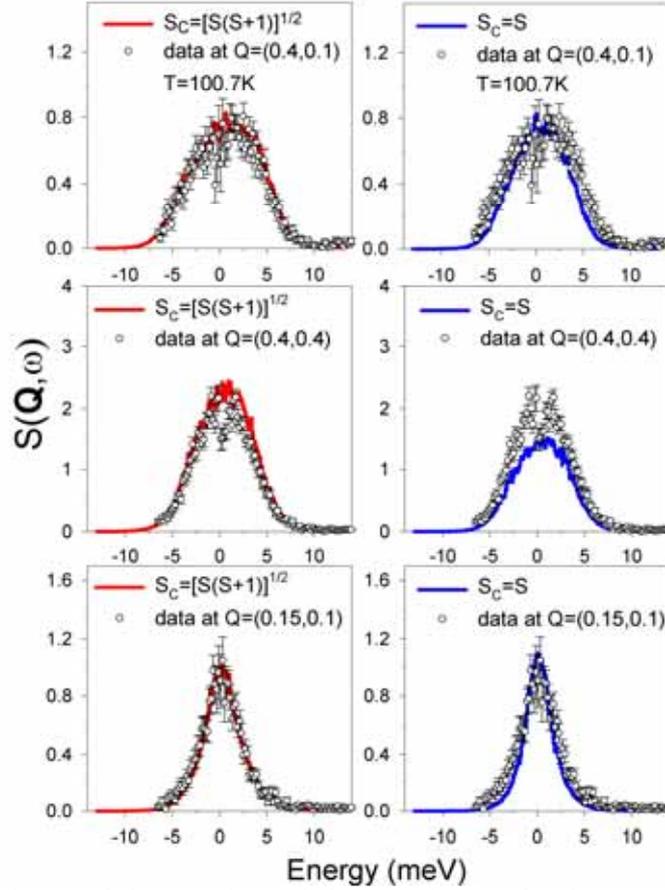

Figure 8 The line shapes of the experimental data at 100.7 K for three wave vectors. They are compared with simulations with parameters from model A on the lhs and with model B on the rhs. The error bars for the measurements are large near zero energy transfer because a large incoherent peak has been subtracted.

**5. Comparison of Experimental and Theoretical Results**
We begin the comparison of the experimental results with the computer simulations at high temperatures 100.7 K. The system is expected to be in a paramagnetic phase and the scattering to be largely quasi-elastic in character, *c.f.* fig. 2c. The neutron scattering measurements, displayed in fig. 8, show the measured profiles and indeed the results are quasi-elastic peaks. Also shown are the results of both simulations: Close to zero energy transfer a large incoherent background has been subtracted from the experimental data so these are less trustworthy than the results with larger absolute energy transfers. The simulations with model A agree very well, with much better agreement than those with the model B parameters. This demonstrates that the system is very well approximated to the classical model at high temperatures using the standard correspondence in spin length *i.e.* $S_C = \sqrt{S(S+1)}$.

# Quantum classical crossover in the spin dynamics of a 2D antiferromagnet

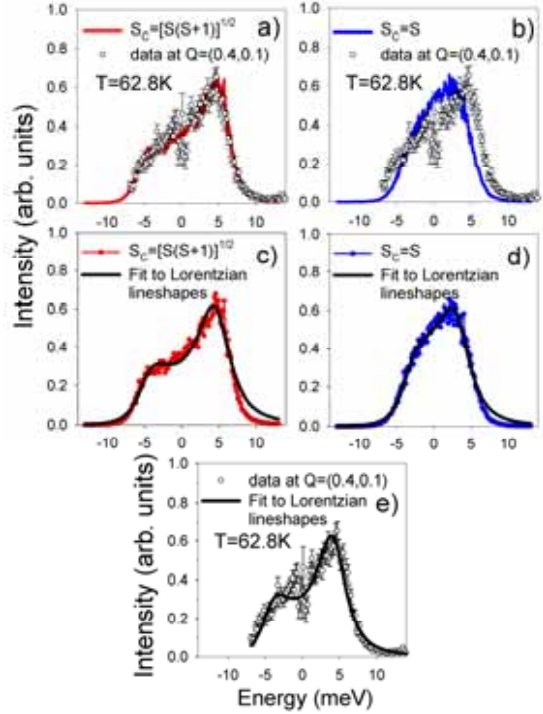

Figure 9 The line shapes of the experimental data at 62.8 K and a wave vector of (0.4,0.1) compared with the results of the simulations a) and b). c) d) and e) show the fitted line shapes for the experiment and for both models used for the simulations.

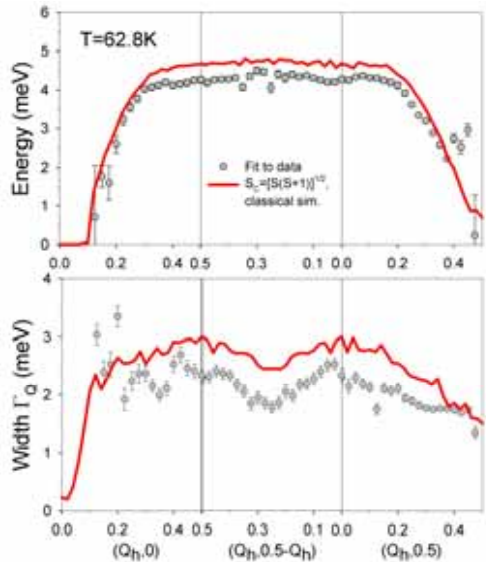

Figure 10 The peak positions and the widths of the excitations as a function of wave vector at 62.8 K The filled circles are the results of the experimental data and the red lines are the results of the simulations with the parameters of model A.

The scattering observed at 62.8K is shown in fig. 9. This is below the Curie-Weiss temperature, and short range correlations are present. In part (a) the experimental data is compared with the simulations using the model A parameters and the agreement is very satisfactory. In contrast the simulation using the model B parameters shown in fig. 9b shows a less satisfactory description of the experimental data because it clearly underestimates the damping of the excitations. The other three parts of fig. 9 show how well the Lorentzian fits provide a description of the results: Figure 9c shows the fit to the Model A, while the fit to

Quantum classical crossover in the spin dynamics of a 2D antiferromagnet

Model B is shown in fig. 9d, and finally fig. 9e shows the fit to the experimental results. From these fits the energies and widths of the excitations are obtained as shown in fig. 10. This provides a more quantitative comparison between the results of experiment and the simulations. For example, the results of model A suggest that the frequency is overestimated slightly by the simulations by about 5% while the line width is also overestimated by about 15%. Some of this discrepancy may be due to the uncertainty of the neutron scattering data around zero energy transfer.

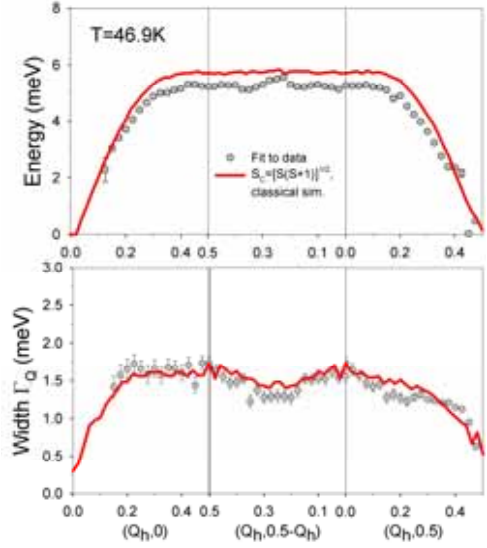

Figure 11  The dispersion relations for the energy and width of the excitations at 46.9 K The filled circles are the results of the experiments and the red lines are the results of the simulations with model A.

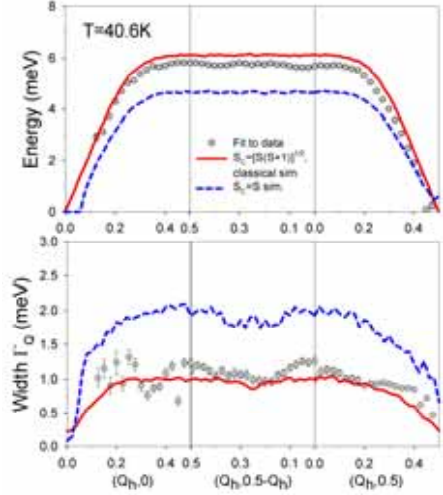

Figure 12 The dispersion relations for the excitation energies and the widths of the excitations at 40.6 K. The filled circles are the results of the experiments, the red solid lines are the simulations with the parameters from model A and the blue dashed lines the simulations with model B.

The simulations using $S_C = \sqrt{S(S+1)}$, Model A, show very good agreement with the temperatures continuing down to 30 K, see figs. 10-14. Therefore this model provides outstanding agreement over a very wide range of temperatures and wavevectors, from paramagnetic down to into the ordered phase. The main changes with decreasing temperature are: As the temperature is reduced to 56.7 K, and 51.6 K the maximum energy of the measured excitations increases to about 4.8 meV and 5.5 meV respectively. The width decreases with decreasing temperature and its maximum value is about 2.1 meV at 56.7 K and 1.8 meV at 51.6 K. At 46.7 K the line-widths have decreased to a maximum of about 1.6

Quantum classical crossover in the spin dynamics of a 2D antiferromagnet meV. Below this temperature the modes are becoming reasonably well defined even though the temperature is still considerably above the Nèel temperature of 38.4 K. Further cooling to 40.6 K brings a rapid decrease in the width of the excitations as shown in fig. 10. The widths have decreased so that the maximum width is now about 1.1 meV and the maximum energy of the excitations has increased to about 5.8 meV. The peaks are well defined excitations especially at the zone boundary. Throughout this temperature range from paramagnetic until just above the ordering temperature, Model B gives very poor agreement with both the energy of the excitations and their line-width, as can be seen in figs. 8 and 12 and is clearly unsuitable.

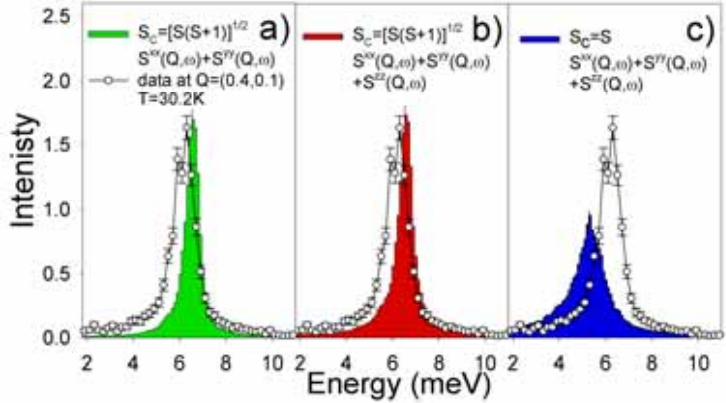

Figure 13 The line shape of the excitations for a temperature of 30.2 K and a wave vector (0.4,0.1) The circles are the experimental data while in panel a) the green histogram is the transverse part of the simulations with parameters from model A, in panel b) the sum of all 3 parts of the scattering, and in panel c).the results of simulations with model B are shown.

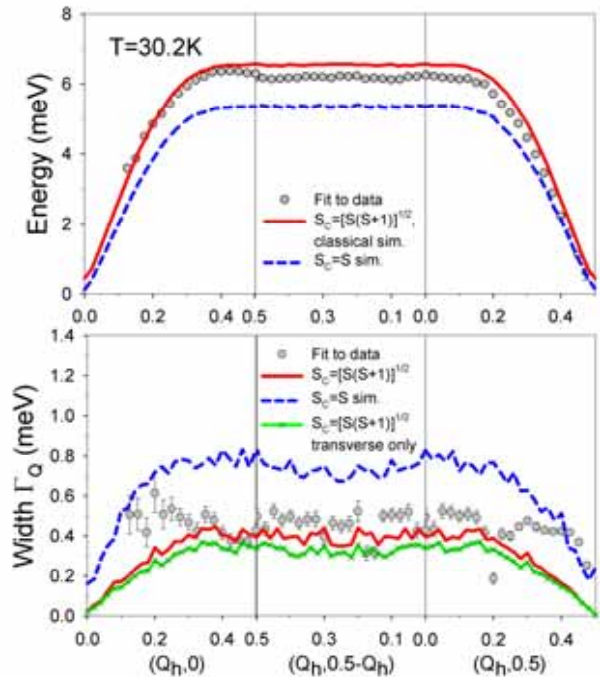

Figure 14 The dispersion relation for the frequencies and widths of the excitations at 30.2 K. The filled circles show the experimental data and the red, green and blue lines are as shown for figure 13.

Below $T_N$, at 35.4K, the energy of the excitations has increased to 6.0 meV with the model A simulation giving energies that are about 5% larger than the experimental results. The

# Quantum classical crossover in the spin dynamics of a 2D antiferromagnet

maximum width is about 0.8 meV which is about 0.1 meV more than the model A simulations. Figure 13 shows the spectra for a temperature of 30.2 K and it is clear that the agreement with the line shape is much better for the simulation with model A parameters than for the simulation with model B parameters. Figure 14 shows the energies and line widths throughout the Brillouin zone and the agreement between the simulations using model A and the experimental measurements is similar to that found at higher temperatures. The line-widths near the antiferromagnetic Bragg peaks are relatively larger than they were at high temperatures and this is possibly due to the difficulty of treating the resolution corrections satisfactorily as discussed above. In contrast, the simulation with model B parameters gives considerably less satisfactory results. At a temperature of 25.5 K the line-width at the zone boundary has decreased to 0.32 meV and at 21.3 K it has further decreased to 0.19 meV. Both of these widths are given very satisfactorily by the simulations with model A parameters. As expected for low temperatures, the estimation of the excitation energies is larger for the simulations with model A parameters than for the experiment and as discussed above this is due to the neglect of the quantum fluctuations that contribute to the measurements. So it is only at temperatures of about 25 K and below that divergences between the $S_C = \sqrt{S(S+1)}$ simulations and real behaviour are seen.

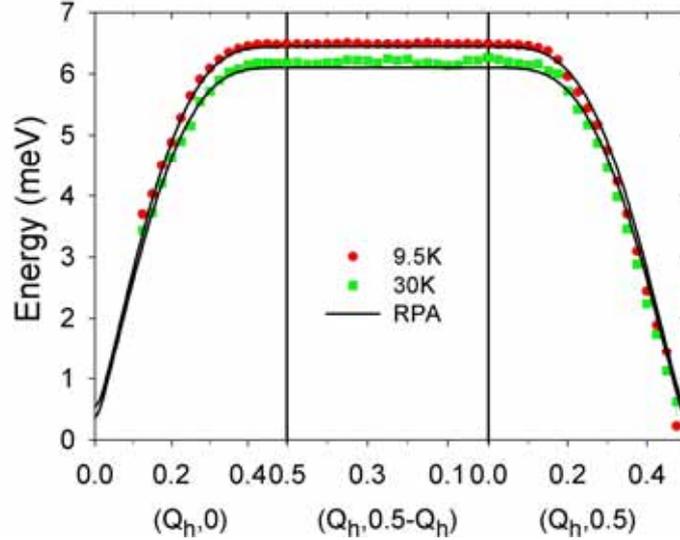

Figure 15 A comparison of the temperature dependence of the dispersion relation as measured by the experiment and as calculated using the random phase approximation using the parameters deduced from low temperature spin wave theory.

The thermal dependence of excitations and line-widths, particularly below $T_N$, are now compared with analytic theories including the quantum behaviour. At low temperatures the excitation frequencies are expected to be described with spin wave theory and indeed the measurements at low temperature have been used to fix the exchange constant. In fig. 15 we show the agreement between the measured energies of the excitations at 9.5 K and at a considerably higher temperature of 30 K with the energies calculated using the random phase approximation, eqns 4 and 5, with the exchange parameters fixed by the low temperature spin wave results. They are all in excellent agreement with the calculations. We conclude that the random phase approximation gives a good description of the excitation energies at least at temperatures well below $T_N$. This theory does not however provide an explanation for the width of the excitations.

Figure 16 summarises the behaviour of the line width at the antiferromagnetic zone boundary and illustrates both the temperature dependence of the excitation energy as well as the temperature dependence of the line-width. The success of the simulations with model A parameters for temperatures above ~35 K is clear and both the energy and the line width are

Quantum classical crossover in the spin dynamics of a 2D antiferromagnet accurately obtained. Below 35 K, the quantum character of the system is evident and the energy predicted by the model A simulation becomes increasingly larger than the measurements at low temperature. A comparison of the line-width is made with the theory of Tyc and Halperin [18] and this is seen to agree relatively well with the zone boundary temperature dependence below about 50K. In contrast the theory of Kopietz [19] agrees less well with the zone boundary line broadening.

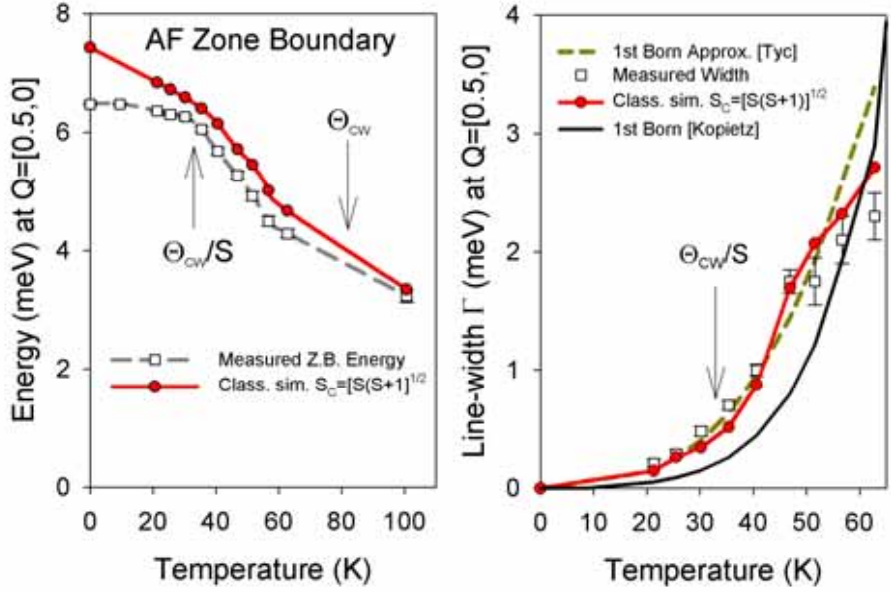

Figure 16 (Left panel) The experimental energy of the zone boundary spin waves with the results of the classical simulations using the parameters from Model A. The experimental point at absolute zero was obtained by extrapolation of the experiments at 9 K. (Right panel) The fitted line widths at the zone boundary are compared to the interacting spin wave theory of Tyč and Halperin [18] computed using the measured temperature renormalisation of the spin wave dispersion shown in left panel. Also shown is the result of the model A classical simulation that agrees very well with the data above $T^* = \theta_{cw}/S$. The analytical form for spin wave damping suggested by Kopietz [19] is also shown and agrees less well with the data.

To clarify the temperature dependence of the line-width further a logarithmic plot is shown in fig. 17. Based on the work of Kopietz [19], we propose a phenomenological form for the temperature dependence of the line broadening at the zone boundary

$$\frac{\Gamma_q}{\omega_{ZB}} = C\left(\frac{k_B T}{\hbar \omega_{ZB}}\right)^\eta F(q) \qquad (10)$$

where the exponent $\eta$ and the constant $F(q)$, are determined from the experimental results. We find that the experimental data is fitted empirically, at least approximately, by a quadratic $\eta = 2$ dependence instead of the cubic $\eta=3$ dependence predicted by Kopietz [19]. This result can be extended at least approximately to all wave vectors. If the line-widths are divided by $T^2$ then all the measured line-widths for the excitations become very similar between 21.3 K and 46.9 K as shown in fig.18. The wave vector dependence can be expected to be expanded around the zone boundary wave vector as an polynomial expansion *i.e.* $F(q) = a_0 + a_2(q-q_{ZB})^2 + a_4(q-q_{ZB})^4...$, and the data over this wide range of temperatures and wave vectors is well described by the model with the parameters $C =0.455$, $\eta =2$, $a_0=1$, $a_2 =-1.35$. Also shown in fig. 18 is the wave vector dependence predicted at one temperature, 21.3 K, by the theories of Tyč and Halperin [18], and by the theory of Kopietz [19]. Both disagree strongly with the wave vector dependence except possibly close to the zone boundary.

Quantum classical crossover in the spin dynamics of a 2D antiferromagnet

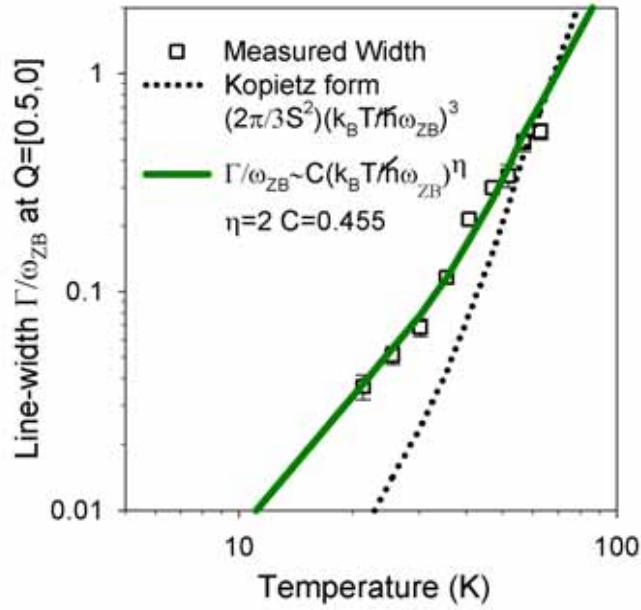

Figure 17 A logarithmic plot of the relative line width versus temperature and a $T^2$ dependence is found to describe the data, i.e. $\Gamma_q/\omega_{ZB} = 0.455 F(q)(k_B T/\hbar\omega_{ZB})^2$ where F(q) and C are dimensionless. The temperature dependence is in clear disagreement with the cubic dependence proposed by Kopietz [19].

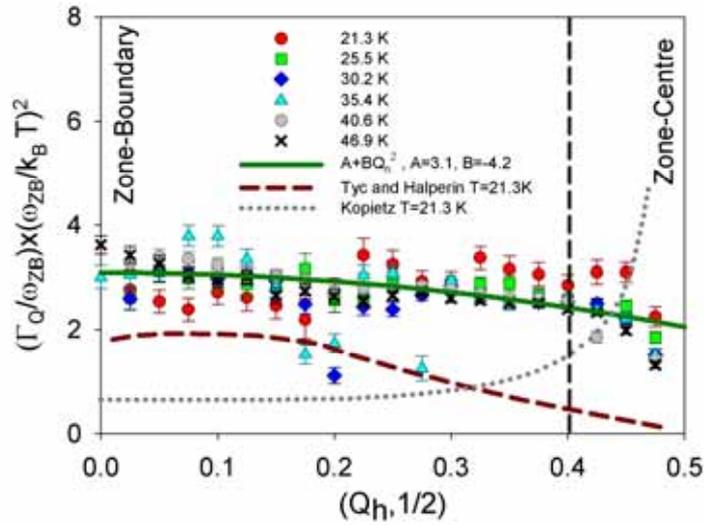

Figure 18 The wave vector dependence of the line widths, rescaled for the quadratic temperature dependence, is displayed. The data from different temperatures roughly overlay each other and show a slight downward trend as the zone centre is approached from the zone boundary. A vertical dashed line is shown to the right of which the magnetic Bragg scattering and critical scattering may influence an accurate determination of the line width. The wave vector dependence given by the Kopietz [19], and Tyč and Halperin [18] theories are also shown when computed for 21.3 K.

The underlying validity of the spin wave theories only extends to regions where $|q|\gg\kappa(T)$ i.e. when the excitation wave vector is larger than the inverse correlation length. Below $T_N$ the system is ordered so all wave vectors are potentially applicable. Above $T_N$ it is anticipated [7] that the dynamics is overdamped (relaxational) for wave vectors $|q|\ll\kappa(T)$. In order to test this expectation we have calculated the wave vector dependence of the ratio of the line width to the excitation energy $\Gamma_q/\omega_q$ for various temperatures. The results are shown in fig. 19 and for the temperatures above $T_N$ the ratio $\Gamma_q/\omega_q$ increases to values above unity near the

Quantum classical crossover in the spin dynamics of a 2D antiferromagnet antiferromagnetic Bragg reflection, i.e. overdamped behaviour for $|q|<<\kappa(T)$. This overdamping for $|q|<<\kappa(T)$ is also given by the results of the simulations.

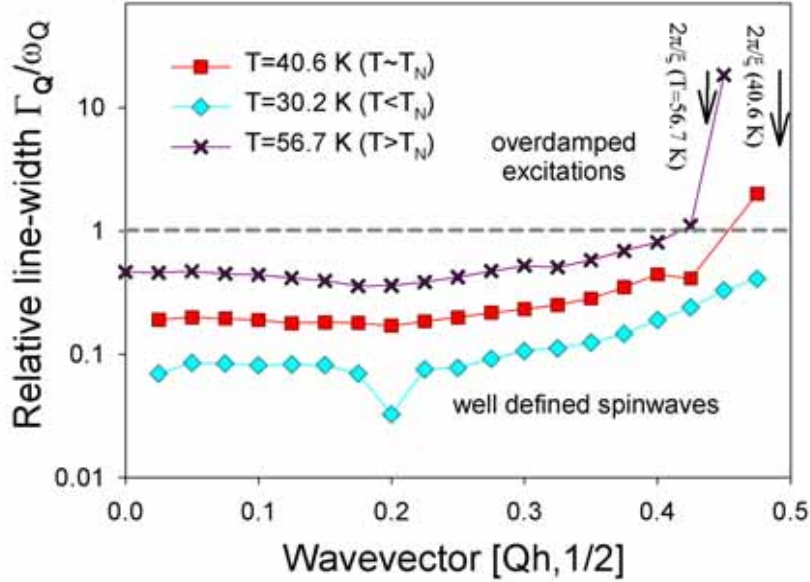

Figure 19. The wave vector dependence of the ratio of line width to excitation energy is shown for three temperatures. For wave vectors near the zone boundary [0,1/2] the spin waves are well defined as $\Gamma_k/\omega_k < 1$. For elevated temperatures relaxational (overdamped) dynamics is observed around the zone centre where $\Gamma_k/\omega_k > 1$. The expectation that this occurs for wave vectors from the antiferromagnetic point $q < \kappa(T)$, is shown and agrees with experimental results

Figure 20 shows the regions of validity for different theories as a function of wave vector and temperature. In the case of the form suggested by Kopietz [19] the line broadening is predicted to be proportional the cube of the temperature, $\Gamma_Q \sim T^3$, and the approximations for the theory are only valid in the range $|q| >> (2\pi/a)[k_B T a/c]^{1/3}$. As is evident from the figure, this range is largely outside the region of experimental measurements which may explain why there is considerable disagreement between this theory and the experimental results. Further experiments should be carried out at lower temperatures across the Brillouin zone to investigate whether the theories of interacting spin waves can be confirmed in the region of $T$ and q for which the theory is supposed to be valid.

For the pure 2D isotropic Heisenberg antiferromagnet CHN propose [7] that dynamic scaling applies for $q << 2\pi/a$ and that a scaling form for the scattering cross section is obeyed i.e. $S(q,\omega) \sim \bar{\omega}_0^{-1} S(q) \Phi(q\xi, \omega/\bar{\omega}_0)$. The phenomenological characterisation of the data in fig. 16 does not however conform to this scaling behaviour. The lack of scaling is perhaps not surprising, as the behaviour of $Rb_2MnF_4$ is far from ideal because $T_N$ is about 1/3$^{rd}$ of the zone boundary energy of the material. Another interesting question is whether hydrodynamic behaviour, as is observed for example in 3D antiferromagnets, will also be observed in 2D antiferromagnets. Hydrodynamics has a characteristic line width dependence of $\Gamma_q \propto Dq^2$ [18].

Within the temperature and wave vector ranges studied no such behaviour has been observed. This is consistent with its absence in the 2D Heisenberg antiferromagnet but more detailed studies with very high energy resolution and at low temperatures should be carried out to establish this more thoroughly asymptotically close to the zone centre.

The Curie-Weiss temperature is the temperature scale on which thermal fluctuations overcome the energy of a spin in an effective field provided by the neighbours

Quantum classical crossover in the spin dynamics of a 2D antiferromagnet

$kT \approx SB_{eff} = \Theta_{CW}$ and the quantization (level spacing) of such a spin in this field is $\Delta E \approx B_{eff}$. As classical dynamics is in general adequate when $kT > \Delta E$ a possible crossover temperature is $T^* = \Theta_{CW}/S$. This criterion agrees well with the data presented above and also the apparent crossovers for the correlation length data shown in Elstner [23] where the question of applicability of quantum or classical descriptions was also considered. Finally, due to the small quantum renormalisation factor $Z_c$ we see this cross-over clearly in $Rb_2MnF_4$.

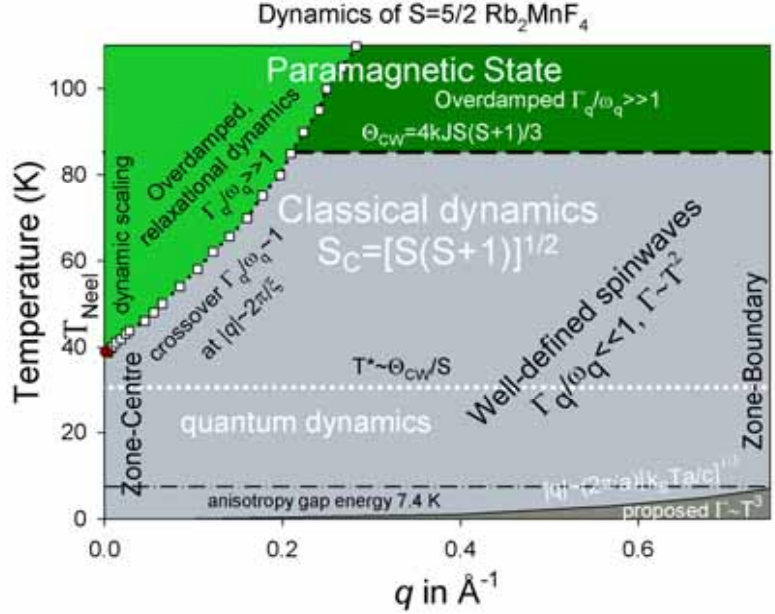

Figure 20 Crossovers in temperature and wave-vector of the dynamical behaviour in $Rb_2MnF_4$. Above the Curie-Weiss transition temperature, $\Theta_{CW} = 87.5$ K [24], the material is paramagnetic and the dynamics is over-damped. Also, for wave-vectors smaller than the inverse correlation length, taken from Lee et al [3], our data and computations also show over-damped behaviour. At small wave vectors, $q \ll q_{ZB}$, dynamical scaling is proposed and has been observed at $k=0$ for $Rb_2MnF_4$ in a field by Christianson et al. [12]. The field counters the small anisotropy and promotes the scaling behaviour. Spin-waves are observed over the rest of the wave-vector and temperature region i.e. $\kappa(T) \ll q$ and $T \ll \Theta_{CW}$ and these show non-universal behaviour and are not described by dynamical scaling. An important finding of this study is that the classical computations show a large region of validity and are accurate above a crossover temperature $T^* \sim 35$ K on all wave vector scales

One interesting consequence of the quantum-classical crossover is in the change of effective spin length from $S$ to $[S(S+1)]^{1/2}$. At low temperatures the mean field effects of the neighbouring spins put particular spins into definite directional states and according to the principles of quantum observation this behaves as a definite spin component of magnitude $S$. Fluctuations of this directional component are the relevant variables. At elevated temperature the directionality of individual spins becomes less defined by the neighbours and approaches the free behaviour of acting as a vector of magnitude $[S(S+1)]^{1/2}$. This then provides a physical picture of the crossover.

6. Summary and Conclusions

The experiments were performed at least in part to test how successful experiments using the time-of-flight spectrometer MAPS would be for measuring the temperature dependence of the excitations throughout the Brillouin zone in a two-dimensional antiferromagnet. The results demonstrate that for spin waves near the zone boundary we could obtain very satisfactory results in a relatively short counting time. Near the zone centre the frequency of the excitations is lower and we could have used improved resolution to measure these spin waves more reliably. This, however, would have entailed a considerable increase in the counting

Quantum classical crossover in the spin dynamics of a 2D antiferromagnet time. It is also difficult to do on MAPS because we used almost the smallest available incident neutron energy but these experiments could have been performed on other neutron time-of-flight spectrometers. This would, however, be partly self-defeating because the counting times would be much longer and the data analysis more complex so that many of the advantages of the time-of-flight technique would be lost.

The second purpose of the experiment was to compare the results with both analytic theory and computer simulations of the scattering. It is not possible to perform a quantum mechanical calculation but we have performed a classical computer simulation by using an enhanced Metropolis algorithm. We have successfully performed a quantitative comparison of the data with the computer simulations. The computing required personal intervention to obtain satisfactory results and a large amount of data was processed much of which does not appear in the final publication. This though was directed at establishing an efficient methodology to apply to quantum magnets, as described in the appendix.

Nevertheless, it is now possible to compare the experimental results and the results of the simulations. The simulations were tested by comparison with the low temperature results that could be evaluated for a classical system and also by comparison with high temperature results. We found that if we used a simulation that took the true exchange values with a classical correspondence of spin length $S_C = \sqrt{S(S+1)}$, although it did not correctly give the lowest temperature excitation energies, we obtained a reasonable description of the data, roughly accurate to better than 5%, over almost all the rest of the temperature range. This was model A of section 5. The deviations can be associated with the failure of the classical model at low temperature. We can therefore consider that the project has been successfully completed and suggest that MAPS makes possible the comprehensive mapping of dynamics over a wide range of wave vectors and temperature not accessible previously.

We have also compared our results with the predictions of analytical theories and found that at low temperatures below the Nèel temperature there was reasonable agreement with experiment both for the random phase approximation theories of the energies of the excitations and for the line width of the zone boundary excitations. The analytic theories did not describe the behaviour at higher temperatures. We have shown that at these temperatures the line width throughout the zone scales as the square of the temperature which is a different scaling from any of the theories [18], [19]. This is unlikely to be due to the fact that $Rb_2MnF_4$ is not an ideal 2DHAFSL as the simulations show that in the temperature ranges concerned the anisotropy has little effect on the excitation energies and line-widths despite ordering. Our results demonstrate, in fig. 20 that the classical model describes the scattering over a very wide range of temperatures and wave vectors which are not adequately described by analytic theories. In order to obtain experimental results which should be compared with the analytic theories the measurements should be made at low temperatures and very small wave vector transfers. This is not possible using conventional neutron scattering techniques but might be possible with inelastic spin echo techniques. These lower temperatures should be particularly interesting as dipolar interactions, which are responsible for the magnetic ordering in $Rb_2MnF_4$ may also modify the excitation line widths here.

Finally, further insight into the success of the classical approach can perhaps be gained by considering Figure 16 again. At low temperatures both the classical and quantum systems are described by spin waves and in both cases the RPA describes the thermal renormalisation of the spin wave dispersion. The classical spin wave dispersion decreases much more rapidly with temperature than the quantum case at low T. The classical RPA was essentially derived by replacing the Bose occupation by an equivalent classical quantity ($n_q \to k_B T / \hbar \omega_q$). This suggests that the difference between the classical and quantum systems is caused by the fact that they obey different statistics, although they obey the same equations of motion and have

Quantum classical crossover in the spin dynamics of a 2D antiferromagnet the same energies. At higher temperatures, the Bose occupation factor tends to the classical limit so the classical and quantum systems converge.

In conclusion, we have measured the frequencies and the line widths of the magnetic excitations in Rb$_2$MnF$_4$ over a very large range of wave vectors and temperatures and the results show surprisingly good agreement with simulations based on a classical model.

**Acknowledgements**

We acknowledge stimulating and useful discussions on this work with Dr Sibel Bayrakci, MPI Stuttgart, Prof. Marshall Luban, Iowa State Unversity, and Profs. John Chalker and Fabian Essler, University of Oxford. We are grateful for the support of the RAL staff in conducting these experiments and to R. J. Christianson and R. L. Leheny for help with the experiment. T.H, R.C and RAC. are grateful to EPSRC for financial support and A.T. is grateful to RAL and Wadham College, Oxford for financial support.

———————————

**Appendix**

**Computational Method**
Classical systems consist of a set of variables $\mathbf{S} = [\zeta_1,...\zeta_\eta]$ which are said to span the phase space of the system. A volume element of the phase space is denoted by $d\mathbf{S} = \prod_i^\eta d\zeta_i$. The interactions are given by the Hamiltonian,

$$H(\zeta_1,...,\zeta_\eta) = H(\mathbf{S}), \qquad (A1)$$

which leads to the equations of motion through the Poisson brackets:

$$\frac{d\zeta_i}{dt} = \{\zeta_i, H(\mathbf{S})\}, \qquad (A2)$$

Evaluation of the equations of motion give the time development of the system $\mathbf{S}(t)$ from the starting configuration $\mathbf{S}(t_0)$. All physical quantities are derived functions of the system variables, $A[\mathbf{S}(t)]$. The ensemble averages for many quantities are time independent quantities and in the canonical ensemble are evaluated as

$$\langle A[\mathbf{S}] \rangle = \frac{1}{Z} \int d\mathbf{S} A[\mathbf{S}] e^{-H[\mathbf{S}]/k_B T} = \frac{\int d\mathbf{S} A[\mathbf{S}] e^{-H[\mathbf{S}]/k_B T}}{\int d\mathbf{S} e^{-H[\mathbf{S}]/k_B T}} \qquad (A3).$$

Time dependent quantities are evaluated in the canonical ensemble as

$$\langle A[\mathbf{S}(t)] \rangle = \frac{1}{Z} \int d\mathbf{S}(t_0) A[\mathbf{S}(t)] e^{-H[\mathbf{S}(t_0)]/k_B T} = \frac{\int d\mathbf{S}(t_0) A[\mathbf{S}(t)] e^{-H[\mathbf{S}(t_0)]/k_B T}}{\int d\mathbf{S}(t_0) e^{-H[\mathbf{S}(t_0)]/k_B T}} \qquad (A4).$$

When considering localised spin systems, the system components are individual components S, and the measure $\mathbf{S} = [S_1^x, S_1^y, S_1^z ... S_\eta^x, S_\eta^y, S_\eta^z]$. The spins are constrained to the classical spin length $(S_i^x)^2 + (S_i^y)^2 + (S_i^z)^2 = (S_C)^2$ so the classical phase space element is $dS = \prod_i dS_i^x dS_i^y dS_i^z \delta\left(S_c^2 - (S_i^x)^2 - (S_i^y)^2 - (S_i^z)^2\right) = \prod_i d\phi_i d\theta_i S_C \sin\theta_i$. The two-spin correlation functions are evaluated as

$$\langle S_{i'}^\beta(t_0) S_i^\alpha(t) \rangle =$$

$$\frac{1}{Z} \int d\mathbf{S}(t_0) S_{i'}^\beta(t_0) S_i^\alpha(t) e^{-H[\mathbf{S}(t_0)]/k_B T} = \frac{\int d\mathbf{S}(t_0) S_{i'}^\beta(t_0) S_i^\alpha(t) e^{-H[\mathbf{S}(t_0)]/k_B T}}{\int d\mathbf{S}(t_0) e^{-H[\mathbf{S}(t_0)]/k_B T}} \quad (A5).$$

The direct numerical evaluation of equation (A4) is not practicable. This is because only configurations very close to the mean energy contribute significantly [25] and a uniform sampling of phase space is too inefficient to be realistically employed [21,26]. Instead importance sampling is used where a set of configurations $[\mathbf{S}_1(t_0),...,\mathbf{S}_M(t_0)]$ is chosen so as not to sample phase space uniformly. The configurations $[\mathbf{S}_1(t_0),...,\mathbf{S}_M(t_0)]$ are chosen such that phase space is sampled with probability density $p[\mathbf{S}(t_0)]$ and the region of configuration space $\Delta \mathbf{S}(t_0)$ is sampled with probability $p[\mathbf{S}(t_0)]\Delta \mathbf{S}(t_0)$. The thermal average (A4) can then be written:

$$\langle A[\mathbf{S}(t)] \rangle = \frac{\sum_{m=1}^M A[\mathbf{S}_m(t)] p^{-1}[\mathbf{S}_m(t_0)] e^{-H[\mathbf{S}_m(t_0)]/k_B T}}{\sum_{m=1}^M p^{-1}[\mathbf{S}_m(t_0)] e^{-H[\mathbf{S}_m(t_0)]/k_B T}} \quad (A6)$$

An algebraic simplification arises if $p[\mathbf{S}(t)] \propto e^{-H[\mathbf{S}(t)]/k_B T}/Z$:

$$\langle A[\mathbf{S}(t)] \rangle = 1/M \sum_{m=1}^M A[\mathbf{S}_m(t)]. \quad (A7)$$

The problem of optimising the distribution of contributing microstates to be peaked around the mean energy is then fixed if the phase space is sampled with $p[\mathbf{S}(t)] = e^{-H[\mathbf{S}(t)]/k_B T}/Z$. This probability density will sample phase space around the mean energy more frequently and therefore more accurately, leading to faster convergence of equation (A7).

To evaluate the dynamical thermal averages the set of configurations $[\mathbf{S}_1(t_0),...,\mathbf{S}_M(t_0)]$ with the distribution $p[\mathbf{S}(t)] = e^{-H[\mathbf{S}(t)]/k_B T}/Z$ is first generated. Then the time evolution $[\mathbf{S}_1(t),...,\mathbf{S}_M(t)]$ from each starting configuration is determined by numerically solving the equations of motion. More specifically, the simulations will generate a value for each spin component at discrete time intervals $S_i^\alpha(t_0), S_i^\alpha(t_1),..., S_i^\alpha(t_N)$. These results can be inserted into (A7) to give $\langle A[\mathbf{S}(t)] \rangle$.

*A.1 Generating the required distributions*
The principle with which to generate the set of configurations $[\mathbf{S}_1(t_0),...,\mathbf{S}_M(t_0)]$ with probability density $p[\mathbf{S}(t)] = e^{-H[\mathbf{S}(t)]/k_B T}/Z$ is to start with an initial configuration $\mathbf{S}^1$ and to modify it in small steps $\mathbf{S}^1 \to \mathbf{S}^2 \to \mathbf{S}^3 \to ...\mathbf{S}^F$ until the final configuration is generated. The modification procedure is to generate new configurations for the system and to replace the existing system with the new one according to some probability $W(\mathbf{S} \to \mathbf{S}')$. It is important that the probability law gives a finite, non-zero probability, for the transition path between any two states of the system [21]. Providing this is satisfied, the condition of detailed balance imposed on $W(\mathbf{S} \to \mathbf{S}')$:

$$\frac{W(\mathbf{S} \to \mathbf{S}')}{W(\mathbf{S}' \to \mathbf{S})} = \exp[H(\mathbf{S}) - H(\mathbf{S}')/k_B T] \quad (A8)$$

ensures that any initial distribution of configurations $[\mathbf{S}_1,...\mathbf{S}_M]$ will approach an equilibrium canonical ensemble $p[\mathbf{S}(t)] = e^{-H[\mathbf{S}(t)]/k_B T}/Z$ as the number of updates $F \to \infty$. Here we employ the Metropolis algorithm [20] where

$$W(\mathbf{S} \to \mathbf{S'}) = \begin{cases} \exp[H(\mathbf{S}) - H(\mathbf{S'})/k_B T] & \text{if } H(\mathbf{S'}) - H(\mathbf{S}) > 0 \\ 1 & \text{otherwise} \end{cases} \quad (A9).$$

The Metropolis algorithm is easily implemented: For each iteration, a random number $r$ is chosen between 0 and 1. The update $\mathbf{S} \to \mathbf{S'}$ is then made if $\exp[H(\mathbf{S}) - H(\mathbf{S'})/k_B T] \geq r$.

*A.2 Equations of motion*

To generate the set of configurations in time $[\mathbf{S}_1(t_0),...,\mathbf{S}_M(t_0)]$ the equations of motion are integrated numerically. The Hamiltonian for $Rb_2MnF_4$ is a 2D nearest-neighbour Heisenberg model with exchange strength $J$ and small z-axis anisotropy $\Delta$:

$$H = J\sum_{i,j(i)}\left(S_i^x S_j^x + S_i^y S_j^y + S_i^z S_j^z(1+\Delta)\right) = \sum_i \mathbf{S}_i \cdot \left\{\sum_{j(i)} J\left(S_j^x, S_j^y, (1+\Delta)S_j^z\right)\right\}$$

$$\equiv \sum_i \mathbf{S}_i \cdot \left\{\mathbf{H}_{eff}\left[\mathbf{S}_{j(i)}(t)\right]\right\} \quad (A10)$$

The spins form two interpenetrating sub-lattices A and B such that spins $i$ are coupled only to their nearest neighbours $j(i)$ which are on the alternate sub-lattice. The effective field on spin $i$ from the neighbouring spins is $\mathbf{H}_{eff}[\mathbf{S}_{j(i)}(t)]$, and causes precession:

$$\frac{d\mathbf{S}_i}{dt} = \mathbf{H}_i^{eff} \times \mathbf{S}_i. \quad (A11)$$

The fourth order Runga-Kutta numerical integration method [27] is used in the simulations:

$$\mathbf{S}_i(t + \delta t) = \mathbf{S}_i(t) + \frac{\delta t}{6}(f_{1i} + 2f_{2i} + 2f_{3i} + f_{4i})$$

$$f_{1i} = \mathbf{H}_{eff}[\mathbf{S}_{j(i)}(t)] \times \mathbf{S}_i(t)$$

$$f_{2i} = \mathbf{H}_{eff}\left[\mathbf{S}_{j(i)}(t) + \frac{\delta t}{2}f_{1j(i)}\right] \times \left(\mathbf{S}_i(t) + \frac{\delta t}{2}f_{1i}\right) \quad (A12).$$

$$f_{3i} = \mathbf{H}_{eff}\left[\mathbf{S}_{j(i)}(t) + \frac{\delta t}{2}f_{2j(i)}\right] \times \left(\mathbf{S}_i(t) + \frac{\delta t}{2}f_{2i}\right)$$

$$f_{4i} = \mathbf{H}_{eff}\left[\mathbf{S}_{j(i)}(t) + \delta t f_{3j(i)}\right] \times (\mathbf{S}_i(t) + \delta t f_{3i})$$

It accumulates an error of only $(\delta t)^5$ and is computationally efficient. The parameters for the simulations are set to run over a total time $t_N$, determined by the required resolution $\delta\omega \approx 2\pi/t_N$. The integration step $\delta t$ is then chosen to give good numerical accuracy up to $t_N$. One test of numerical accuracy is conservation of spin length, $|\mathbf{S}_i|$. The time interval for which data is stored is determined by the largest energy scale of the system $\omega_{max} \approx 2\pi/\Delta t$.

*A.3 Dynamical correlations*

The dynamical quantities of interest for neutron scattering are the Fourier transformed dynamical two-spin correlation functions, defined for the classical system:

$$S^{\alpha\beta}(\mathbf{Q},\omega) = \frac{1}{2\pi N}\sum_{i,i'} e^{i\mathbf{Q}\cdot(\mathbf{R}_i - \mathbf{R}_{i'})} \int_{-\infty}^{\infty} e^{-i\omega t} \langle S_i^\alpha(t_0) S_{i'}^\beta(t_0+t)\rangle dt \quad (A13).$$

The simulations produce a discretized set of spin configurations up to time $t_N$. First, the consequencies of discretisation are discussed. This is followed by a method that we introduce

Quantum classical crossover in the spin dynamics of a 2D antiferromagnet to calculate the thermodynamic average of dynamical quantities that saves considerably on computer time and can be applied to many systems.

From equation A13 above, it can be seen that evaluating $S(\mathbf{Q},\omega)$ involves the integral

$$\int_{-\infty}^{\infty} e^{-i\omega t} S_i^\alpha(t_0) S_{i'}^\beta(t_0+t) dt . \qquad (A14)$$

With a simulation output that is discrete in time this becomes

$$\sum_{n=-\infty}^{\infty} S_i^\alpha(t_0) S_{i'}^\beta(t_0+n\Delta t) e^{i\omega n\Delta t} \Delta t \qquad (A15)$$

which is now periodic, since it is unchanged when $\omega \to \omega + 2\pi/\Delta t$. Therefore sampling at discrete time values sets a maximum value of $\omega$, given by $\omega_{max} = 2\pi/\Delta t$, for which the value of $S(\mathbf{Q},\omega)$ can be extracted. When the finite time cut-off for the simulations is introduced and the convolution theorem applied

$$\sum_{n=-\infty}^{\infty} S_i^\alpha(t_0) S_{i'}^\beta(t_0+n\Delta t) e^{i\omega n\Delta t} \Delta t$$

$$\approx \sum_{n=0}^{N} S_i^\alpha(t_0) S_{i'}^\beta(t_0+n\Delta t) e^{i\omega n\Delta t} \Delta t = \sum_{n=-\infty}^{\infty} S_i^\alpha(t_0) S_{i'}^\beta(t_0+t_n) \Theta(t_n) \Theta(t_N-t_n) e^{-i\omega t_n} \Delta t \qquad (A16).$$

$$= \frac{1}{N} \sum_{\omega'} e^{i(\omega-\omega')(1-N)\Delta t/2} \sum_{n=-\infty}^{\infty} S_i^\alpha(t_0) S_{i'}^\beta(t_0+n\Delta t) e^{i\omega' n\Delta t} \frac{\sin((\omega-\omega')N\Delta t/2)}{\sin((\omega-\omega')\Delta t/2)} \Delta t$$

For large but finite $N$, $\frac{\sin((\omega-\omega')N\Delta t/2)}{\sin((\omega-\omega')\Delta t/2)}$ is highly peaked up at $\omega = \omega'$, with half width $2\pi/t_N$. This means that any feature on the scale $\delta\omega < 2\pi/t_N$ cannot be resolved and so must be set appropriately. Finite size effects in A16 leads to the correlation functions having a small imaginary component and going negative for small numbers of samplings.

A method of saving an enormous amount of computational time, which can be applied to a wide variety of realistic systems when evaluating $S(\mathbf{Q},\omega)$ from the simulation results, will now be outlined. So far, in order to evaluate a quantity such as the spin correlation function, we must generate a set of initial configurations $[\mathbf{S}_1(t_0),...,\mathbf{S}_M(t_0)]$ with the probability density $p[\mathbf{S}(t)] = e^{-H[\mathbf{S}(t)]/k_BT}/Z$. After evaluating the temporal development, physical quantities for each configuration are evaluated. For example the spin-pair correlation function evaluated from one configuration is given by $\sum_{n=0}^{N} S_i^\alpha(t_0) S_{i'}^\beta(t_n) e^{-i\omega t_n}$. With the results, equation (A7) can be used to calculate the thermodynamic average. This method will be referred to as taking the *thermal average*. Typically several hundred starting configurations $[\mathbf{S}_1(t_0),...,\mathbf{S}_M(t_0)]$ are required when using this method.

An alternative method of evaluating the thermodynamic average is by taking, what will be referred to as, the *time average*. With this method, rather than generating several hundred starting configurations $[\mathbf{S}_1(t_0),...,\mathbf{S}_M(t_0)]$ and solving the equations of motion with each starting configuration, we instead use a much smaller number of representative configurations. Consider just one starting configuration $\mathbf{S}_1(t_0)$ selected with the probability $p[\mathbf{S}(t)] = e^{-H[\mathbf{S}(t)]/k_BT}/Z$. For a reasonably large system the likelihood of this being representative is overwhelming. The simulation output is the time development of this single state:

$$\mathbf{S}_1(t_0) \quad \mathbf{S}_1(t_2) \quad \mathbf{S}_1(t_3) \quad \mathbf{S}_1(t_4)... \quad \mathbf{S}_1(t_N).$$

Quantum classical crossover in the spin dynamics of a 2D antiferromagnet

Now the set of states $[\mathbf{S}_1(t_0),...,\mathbf{S}_1(t_N)]$ are all possible configurations of the system, with the same energy $E$. The *time average* then uses the states $[\mathbf{S}_1(t_0),...,\mathbf{S}_1(t_N)]$ as starting configurations. The motivation for this is that generating the $M$ independent states $[\mathbf{S}_1(t_0),...,\mathbf{S}_M(t_0)]$ is by far the most computationally intensive stage. Reducing this to the minimum of starting configurations and maximising the time averaging is optimal. Using $[\mathbf{S}_1(t_0),...,\mathbf{S}_1(t_N)]$ $N+1$ starting configurations can be generated in a small fraction of the time. Also, the equations of motion only have to be solved once to generate the time dependence of effectively $N+1$ configurations:

$$\begin{array}{ccccc} \mathbf{S}_1(t_0) & \mathbf{S}_1(t_1) & \mathbf{S}_1(t_2) & \mathbf{S}_1(t_3)... & \mathbf{S}_1(t_N) \\ \mathbf{S}_1(t_{-1}) & \mathbf{S}_1(t_0) & \mathbf{S}_1(t_1) & \mathbf{S}_1(t_2)... & \mathbf{S}_1(t_{N-1}) \\ \vdots & & & & \vdots \\ \mathbf{S}_1(t_{-N}) & \mathbf{S}_1(t_{-N+1}) & \mathbf{S}_1(t_{-N+2}) & \mathbf{S}_1(t_{-N+3}) & \mathbf{S}_1(t_0) \end{array}$$

This set, of the time development of $N+1$ configurations, can now be inserted into A6 to get the thermodynamic average. This is equivalent to taking the thermodynamic average in the microcanonical ensemble.

There is one important catch which needs to be considered when taking a *time average*. In order for the method to work the system needs to be ergodic. This means that starting from an initial condition $\mathbf{S}_1(t_0)$ with energy $E$, the system will evolve in such a way so that it passes through each region of phase space with energy $E$ for a time proportional to the fractional volume of that region in phase space [28]. Although this property is difficult to prove analytically, it is intuitively expected to hold for sufficiently chaotic systems. The method also places another constraint on $t_N$, which must now be large enough so that the system has representatively covered a significant amount of phase space as it evolves in time.

Using the time average method, the thermodynamic average of the spin-pair correlation function is given by:

$$\left\langle S_i^\alpha(t_0) S_{i'}^\beta(t_0 + t_n) \right\rangle = \\ \frac{\Delta t}{t_N}\left[ S_i^\alpha(t_0)S_{i'}^\beta(t_0+t_n) + S_i^\alpha(t_1)S_{i'}^\beta(t_1+t_n) + S_i^\alpha(t_2)S_{i'}^\beta(t_2+t_n)... \right] \quad (A17)$$

where the number of terms used in the series is $t_N / \Delta t$.

Another simplification arises when the temporal Fourier transforms are taken:

$$\sum_{n=-\infty}^{n=\infty} \left\langle S_i^\alpha(t_0) S_{i'}^\beta(t_n) \right\rangle e^{-i\omega t_n}$$

$$\approx \frac{(\Delta t)^2}{t_N} \times \left[ S_i^\alpha(t_0)\sum_{n=0}^{N} S_{i'}^\beta(t_n)e^{-i\omega t_n} + S_i^\alpha(t_1)\sum_{n=-1}^{N-1} S_{i'}^\beta(t_n+t_1)e^{-i\omega t_n} + ... \right]$$

$$= \frac{(\Delta t)^2}{t_N} \times \left[ S_i^\alpha(t_0)e^{i\omega t_0}\sum_{n=0}^{N} S_{i'}^\beta(t_n)e^{-i\omega(t_n+t_0)} + S_i^\alpha(t_1)e^{i\omega t_1}\sum_{n=-1}^{N-1} S_{i'}^\beta(t_n+t_1)e^{-i\omega(t_n+t_1)} + ... \right] \quad (A18)$$

$$= \frac{(\Delta t)^2}{t_N} \times \sum_{m=0}^{N} S_i^\alpha(t_m)e^{-i\omega t_m} \sum_{n=0}^{N} S_{i'}^\beta(t_n)e^{-i\omega t_n}$$

This is another important time saving step, because the process of taking $N+1$ Fourier transforms is now reduced to taking only 2. Note that in taking the summations here, the evaluation of both spin components was taken over the same domain of times *i.e.* $t_0$ to $t_N$. The procedure here has a clear physical interpretation. It is the time averaged correlation function measured where a shutter has exposed the "sample" for measurement i.e. to the beam and detectors for a period $t_0$ to $t_N$.

Quantum classical crossover in the spin dynamics of a 2D antiferromagnet

Using the equations above, the Fourier transformed two-spin correlation function, $S^{\alpha\alpha}(\mathbf{Q},\omega)$, can now be written:

$$S^{\alpha\alpha}(\mathbf{Q},\omega) = \frac{1}{2\pi N}\frac{(\Delta t)^2}{t_N}\left|\sum_i\sum_{n=0}^{N} S_i^{\alpha}(t_n)e^{i\mathbf{Q}\cdot\mathbf{R}_i}e^{i\omega t_n}\right|^2. \qquad (A19)$$

This conveniently takes care of the phase factor in equation A16 which was the result of a finite time cut-off. Another convenient feature of equation A19 is that it ensures that $S^{\alpha\alpha}(\mathbf{Q},\omega)$ is always real and positive. These are fundamental properties of $S^{\alpha\alpha}(\mathbf{Q},\omega)$, since it is related to a scattering cross-section. These properties were not enforced when taking the *thermal average*. This is because the quantity $\sum_{n=0}^{N} S_i^{\alpha}(t_0)S_{i'}^{\beta}(t_n)e^{i\omega t_n}$ evaluated with one starting configuration will in general be complex and the real part may be negative. By taking the *themal average* over hundreds of starting configurations $[\mathbf{S}_1(t_0),...,\mathbf{S}_M(t_0)]$, the complex and negative parts will tend to zero, but will never actually be completely zero.

*A.4 Implementation*

The Metropolis algorithm was used to prepare configurations. The couplings are nearest neighbour so for any two next-nearest neighbour spins the process of moving one spin and making the Metropolis choice, and then moving the other and applying the Metropolis choice is identical to the process of moving both spins simultaneously and (simultaneously) applying to each spin the appropriate Metropolis choice. Therefore, by using a vectorised routine, the method can be applied to all the spins on the same sublattice.

In order to improve convergence an over-relaxation algorithm was used [29] in addition to the Metropolis algorithm. In an over relaxation step, a new configuration is selected that is as far from the previous step in phase space, without changing the energy of the system. This helps to cover phase space more effectively than spins changed at random. However, as the over relaxation step does not change the energy of the system it can only cover a restricted part of the phase space and so cannot be used on its own.

Implementation of the over relaxation algorithm in classical Heisenberg systems is straight forward [30]. One selects the spin site *i* to be altered. For the Hamiltonian, the energy of that spin is then not changed if it is rotated about the effective field axis **N** parallel to $\mathbf{H}_{eff}[\mathbf{S}_{j(i)}(t)]$. The over relaxation step then consists of rotating the spin $S_i$ by 180° about **N**. This can easily be carried out for all the spins on one sub-lattice as for the Metropolis routine. The Metropolis and over-relaxation algorithms can then be used in any proportions. For the simulations they were used equally.

For starting configurations, Néel order and a completely random paramagnetic state were both used so as to serve as a check. Systems of 100x100 with periodic boundary conditions were used and $1\times 10^5$ cooling steps typically used to generate the equilibriated starting state. Dynamics calculations were undertaken using the Runge Kutta method with between $10^4$ and $10^5$ time steps being recorded. Remarkably, as few as 10 starting configurations were required to gain reliable simulations when the correlation functions were computed using the time averaging equation A19.

Using a 2 GHz laptop with 1Gb of RAM and codes written in Matlab®, which is at least a factor of three slower than compiled Fortran, the Metropolis and over relaxation algorithms were run with $10^4$ cooling and $10^4$ equilibration steps in 17 minutes of cpu time. Dynamics and correlation functions of the 100x100 spin system were calculated over 2100 time steps in 100 seconds. The approach then is efficient and practical for calculating classical correlation functions for comparison to a host of spin systems.

Quantum classical crossover in the spin dynamics of a 2D antiferromagnet